\documentclass[preprint]{aastex} 
\newcommand{\gsim}{\mbox{$\stackrel {>}{_{\sim}}$}} 
\newcommand{\lsim}{\mbox{$\stackrel {<}{_{\sim}}$}} 
 
\slugcomment{Accepted to the Astronomical Journal} 
 
\shorttitle{CO(2-1) in NGC 5253} 
\shortauthors{Meier et al.} 
 
\begin{document}

\title{Molecular Gas and the Young Starburst in NGC 5253, Revisited} 
 
\author{David S. Meier\altaffilmark{1}, \& Jean L. Turner}  
\affil{Department of Physics and Astronomy, UCLA, Los Angeles, CA 90095--1562}
\email{meierd@astro.uiuc.edu;turner@astro.ucla.edu} 
\and 
\author{Sara C. Beck}  
\affil{Department of Physics and Astronomy, Tel Aviv University,  
69978 Ramat Aviv, Israel} 
\email{sara@wise.tau.ac.il} 

\altaffiltext{1}{present address: Department of Astronomy, University
of Illinois, Urbana-Champaign, 1002 W. Green St., Urbana, IL 61801}

\begin{abstract} 
 
We report the detection of CO(2-1) and 3.1 mm and 1.3 mm continuum
emission towards the extremely young starburst in NGC 5253, with data
taken from the Owens Valley Millimeter Array.  Faint CO emission
originates in five molecular clouds distributed along the prominent
dust lane seen in visual images.  With the gas, the morphology of NGC
5253 looks much like a dwarf elliptical version of the ``dust-lane
ellipticals'' or ``polar-ring'' class of galaxies.  The molecular gas
resides in GMCs well away from the radio-IR super-star
cluster/supernebula seen in the radio and infrared.  The millimeter
continuum data confirm that the 2 cm flux from the supernebula is
optically thick; the Lyman continuum rate derived from the 1.3 mm
continuum is N$_{Lyc}~\sim 6 \times 10^{52} s^{-1}$ for the central
$\sim 20^{''}$. CO may underestimate the true molecular column
density, as expected for a low metallicity system, although there are
regions along the dust lane that appear to have near-Galactic
conversion factors.  We estimate a total molecular gas mass of
$M_{H_{2}}\lsim 10^{7}~M_{\odot}$.  The molecular gas in the dust lane
is falling into the galaxy, supporting an accretion hypothesis.  The
dust lane gas cannot therefore be causally associated with the current
burst of star formation.  A relatively small amount, $M_{H_{2}}\lsim
5\times10^{5}~M_{\odot}$, of molecular gas is associated with the
current starburst.  We estimate a star formation efficiency of at
least 25 \% and more likely $\sim$75 \%, consistent with the formation
of a bound cluster.  Despite the extreme youth of the starburst, the
specific trigger of the starburst remains elusive, although the infall
of gas in the dust lane suggests that there is more star formation to
come in NGC~5253.
 
\end{abstract} 
 
\section{Introduction} 
 
Star formation in dwarf galaxies can be as intense or more so, in
terms of stars per unit mass, than those seen in large spirals, but
occurs without the familiar triggering mechanisms provided by
dynamical features such as spiral arms or bars \citep[eg.,][]{B00}.
High resolution observations in the optical/UV and NIR indicate that
the dominant mode of star formation in nearby starbursts is through
massive clusters \citep*[eg., ][]{WS95,OGH94}.  Observational and
theoretical evidence suggests that tidal interaction/merger events may
be an important triggering mechanism for \citep*[see][for a recent
review]{W01}, but the nature of the triggering in isolated dwarfs is
not well understood. Because their metallicities are low, dwarf
galaxies may present galactic conditions more like those seen during
the first epoch of star and cluster formation in the early
universe. NGC 5253 is an excellent example of such a galaxy.

NGC 5253 is a nearby \citep[3.8 Mpc;][Sakai 2001, private comm.]
{CP89,SSLSTPM95,Get00}, dwarf starburst galaxy that may be a companion
to M83.  Its star formation is particularly extreme and dramatic.  The
central region of NGC 5253 contains several ``super-star clusters'' of
$M_{V}\sim$ -10 to -11
\citep*[]{MHLKRG95,G96,BTHLK96,CMBGKLS97,TBH00}, with L$_{IR}\sim ~
10^{9}$ L$_{\odot}$, or two orders of magnitude larger than any
individual Galactic star forming regions \citep[eg.,][]{GTB01}.
Strong Br$\gamma$ emission, weak CO bandhead, bright Wolf-Rayet
features and a radio spectrum consistent with being entirely thermal
demonstrate that this burst of super star cluster (SSC) formation must
be very young \citep*[][]{CTM86,RLW88,C91,BTHLK96,THB98}.  Models of
the stellar population based on optical colors indicate that the
youngest optical/UV clusters have ages of $\sim$2.5 Myr up to $\sim$50
Myrs \citep[eg.;][]{CMBGKLS97,TCLH01}.  Radio and infrared
observations find an even younger, more obscured super star-cluster
\citep*[][]{THB98,TBH00,GTB01}.  The starburst in NGC 5253 is
extremely young, possibly the youngest yet observed
\citep*[][]{RLW88,TBH00, GTB01}.  The molecular gas distribution from
which this young burst originated should still be evident.

\citet*[][]{TBH97} observed NGC 5253 in CO(1-0) with the Owens Valley
Millimeter Interferometer (OVRO), and managed a tentative
(5-7$\sigma$) detection of CO with $\sim 11^{''}~ - ~15^{''}$
resolution.  Although the CO detection is tentative it lines up with
the visible dust lane.  The CO emission is extremely weak considering
the amount of star formation present.  It is two orders of magnitude
weaker than its neighboring spiral galaxy, M 83, which has a starburst
of similar luminosity \citep[Turner, Hurt \& Kenny 2002, in
preparation;][]{CTBHM02}.  The ratio of free-free luminosity to
inferred molecular mass in NGC~5253 is 
$L_{OB}/M_{H_{2}}~\gsim~ 4000$, as compared
to $\sim$10 - 20 in nearby, actively star-forming spirals.  
\citet[][]{TBH97} argued, on the basis of
 the CO(1-0) morphology, that a cloud of gas, possibly with
ultra-low metallicity, is accreting onto the galaxy and has triggered the
starburst.

To confirm the CO detection and to further investigate the nature of
the molecular gas and its connection to the starburst in NGC 5253, we
have obtain higher sensitivity and resolution observations of the
J=2-1 transition of CO with OVRO.
 
\section{Observations and Data Reduction} 
 
Aperture synthesis observations of the CO(2-1) transition (230.538
GHz) were made with the Owens Valley Radio Observatory (OVRO)
Millimeter Interferometer between 16 November 1996 and 21 December
1997 (Table 1).  The interferometer consists of six 10.4m antennas
with cooled SIS receivers \citep[][]{OVRO91,OVRO94}.  System
temperatures (single sideband) ranged from 700 - 2200 K.  A 64 channel,
2 MHz filterbank was used giving a velocity resolution of 2.6 km
s$^{-1}$, which corresponds to an overall bandwidth of 166 km
s$^{-1}$.  To increase the signal-to-noise, the data were smoothed to
4 MHz resolution.  Channel 32.5 is centered at $v_{LSR}$= 410 km
s$^{-1}$.  Separate simultaneous 1 GHz bandwidth continuum
observations at 97 GHz (3.1 mm) and 233 GHz (1.3 mm) were obtained.
Two pointings with phase centers of $\alpha_{1}$(B1950) = 13:37:04.80;
$\delta_{1}$(B1950) = -31:22:58.0, $\alpha_{2}$(B1950) = 13:37:05.60;
$\delta_{2}$(B1950) = -31:23:22.0, were observed to cover the nuclear
starburst and dust lane.  Phase calibration was done by observing
quasars, 1244-255 and 1334-127, every 20-30 minutes.  For absolute
flux calibration we used Neptune as primary flux calibrator and 3C273
as a secondary flux calibrator.  Absolute fluxes are good to
$\sim$25\% - 30\% for the 1 mm data and 15\% for the 3 mm data, with
uncertainties due to brightness variations in the calibrators and
elevation effects at this low declination. 
 
The dataset was mosaicked using MIRIAD software.  The maps are
robustly-weighted and corrected for primary-beam attenuation.  Since
the maps are mosaics, the noise level varies across the maps.
Reported noise levels for this paper are those measured from line free
regions of the map half-way between the map center and the edge (FWHM
of the primary beam).  The noise level is a bit lower than this in the
center and somewhat higher toward the edges of the map.  All further
data reduction was done using the NRAO AIPS package.  To generate an
integrated intensity map (I$_{CO}$) velocities were integrated from
350 km s$^{-1}$ to 450 km s$^{-1}$.  Only emission greater than
1$\sigma$ in the channel maps was included in I$_{CO}$ to minimize the
effect of summing noise.  For the faint, narrow lines in NGC 5253, it
is possible that this prceedure might clip out a small amount of real
emission.  However, we tested different clip levels and found that for
different clip levels neither the morphology on the intensity varied
more then the given noise level.
 
Tracks in Equatorial (E) and Low (L) configurations were used to make
up the dataset.  In the resolution setting configuration (E) one
antenna along the E-W arm of the array was down, making this
configuration almost entirely N-S.  This gives the 1 mm beam an E-W
extension not typically seen for such low declination sources.  The
(u,v) coverage for these observations implies that emission on scales
larger than 20$^{''}$ is resolved out.  Given the small field of view
at 230 GHz and the differences in pointing centers between our data
and single-dish CO(2-1) data \citep[]{MTCB01}, it is difficult to
quantify exactly how much flux is resolved out.  A single-dish peak
main-beam temperature of 34 mK (1.3 Jy for the 30$^{''}$ beam) was
detected for CO(2-1).  We estimate the flux resolved out in two ways.
Firstly, the flux in the channel maps were summed over the single-dish
beam area.  A peak flux of 1.4 Jy was obtained, consistent with
detecting all the emission.  Secondly, we convolved the observed maps
to a resolution of 30$^{''}$ and compared it to the single-dish data.
Only a 0.65 Jy peak was detected, in this method, suggesting
$\sim$50\% of the flux is resolved out.  Therefore, we estimate that
towards the localized peaks most of the flux is detected, but that we
probably resolve out some widespread diffuse emission.
 
\section{Results} 
\subsection{The Molecular Gas Morphology} 

Figure 1 displays the CO(2-1) integrated intensity overlaid on the HST
V-band and H$\alpha$ images of NGC 5253 \citep[][]{CMBGKLS97}.
Molecular gas is largely confined to the prominent dust lane that
stretches across the minor axis of the galaxy, confirming the results
of \citet[][]{TBH97}.  In the field-of-view of these observations, no
evidence is seen for a ``disk'' of CO emission oriented along the
major axis of the galaxy, though it is possible that weak extended
emission has been resolved out if it is smooth on $\sim$20$^{''}$
scales.  Very little molecular gas is detected towards the central
super-star cluster/radio ``supernebula."  The rms noise level of 0.6 K
km s$^{-1}$ corresponds to a column density of $N(H_{2})\simeq 3.5
\times 10^{20} ~cm^{-2}$ for the Galactic conversion factor, or
$A_{V}\sim 0.4$, averaged over the 90$\times$180 pc beam.  Only the
faintest cloud, GMC D, is within 200 pc of the central cluster.  For
the first time molecular gas is tentatively detected on the northwest
side of the galaxy.  It lies on the same line traced by the dust lane.
The CO along the dust lane seen prominently to the southeast appears
to cross the galaxy and continue to the northwest.  The molecular gas
also lines up with the prominent H$\alpha$ and [OIII] filament
\citep[][]{G81,CMBGKLS97}.  NGC 5253 is, though much smaller,
reminiscent of the ``dust lane ellipticals'' or ``polar-ring'' classes
of galaxies \citep*[eg.,][]{SG93,GSS97}.
 
CO(2-1) emission resolves into five clouds along the dust lane
(labeled A - E).  Clouds D \& E are tentatively detected, being only
$\sim3-5\sigma$ in intensity, with cloud D being the faintest.  Clouds
are indentified as any spatially and spectrally localized clump that
has an intensity greater than 3$\sigma$ in two adjacent channels.
Each cloud has been fit with an elliptical gaussian to determine its
size and location.  A box containing all the emission from the cloud
was then summed to produce a spectrum, which was in turn fitted with a
gaussian to obtain linewidths and velocity centroids (Table 2).  Only
cloud D is slightly resolved.  Taking the beam minor axis as an upper
limit, the remaining clouds have sizes (diameters) less than 90 pc.
Typically giant molecular clouds (GMCs) in less-active dwarf galaxies
have sizes $\lsim$10 - 70 pc and linewidths of $\sim$3 - 12 km
s$^{-1}$ \citep*[eg.][]{RLB93,W94,W95,THKG99,WTHSM01,MTB01}.  In the
region associated with cloud C, the dust lane is particularly well
pronounced against the background galaxy (Figure 1).  If we assume
that the molecular gas clouds follow the area of optical extinction, a
size of $\sim3^{''}$ or $\sim$55 pc is estimated.  This suggests that
the upper limits listed in Table 2 for the sizes and virial masses are
about a factor of two larger than their true values.  Linewidths for
the clouds along the central dust lane are narrowest (8 - 15 km
s$^{-1}$).  The upper limits of the cloud sizes and linewidths in NGC
5253 are, therefore, consistent with being GMCs, and henceforth will
be referred to as such, although we cannot determine from the current
data whether to identify them as self-gravitating objects (\S 3.2).
 
\subsection{The Molecular Cloud Masses in NGC 5253} 

The estimation of molecular gas masses in dwarf galaxies is difficult,
but important for determining the strength and efficiency of star
formation.  In environments with low metallicities and high radiation
fields, it is predicted that CO ceases to be a reliable tracer of
molecular hydrogen \citep[][]{MB88,E89,S96}.  We can hope to converge
on the true molecular mass of NGC 5253 by taking virial masses as
upper limits to the mass and masses from the Galactic conversion
factor as likely lower limits to the gas mass. We describe these
methods below.

The first to consider is the use of the standard conversion factor:
\begin{equation} 
M_{CO}~=~5.11\times 10^{4}~ (M_{\odot}) \left({X^{N5253}_{CO}} \over
{X_{CO}}\right) \left(\frac{D}{3.8 Mpc}\right)^{2}~{S_{CO}\over
R_{21/10}},
\end{equation} 
where $X^{N5253}_{CO}$ is the conversion factor applicable to NGC
5253, the Galactic conversion factor $X_{CO}$ = $2.0\times
10^{20}~cm^{-2}(K km s^{-1})^{-1}$ \citep[][]{Set88,Het97}, $S_{CO}$
is the CO(2-1) flux in Jy km s$^{-1}$, and $R_{21/10}$ is the
CO(2-1)/CO(1-0) line ratio \citep[eg.,][]{PW98,MTCB01}.  In all
further calculations, $R_{21/10}$ is assumed to be unity, consistent
with the observed single-dish value \citep[][]{MTCB01}.  For
$X^{N5253}_{CO}/X_{CO}~=~1$, we find that the masses of the GMCs range
from $4-9\times 10^{5}~M_{\odot}$ (Table 2).  In low metallicity
systems, CO emission is generally found to be underluminous for a
given gas mass with respect to molecular gas in the Galaxy \citep{I86,
SSLH92, VH95, W95, AST96,TKS98}.  Hence the assumption
$X^{N5253}_{CO}/X_{CO}~=~1$ should give a lower limit to the true
molecular gas present.
 
A second method makes use of the assumption that the molecular clouds
are in virial equilibrium, so that the constraints on clouds size and
linewidth can be related to the total gravitational mass present
\citep*[eg.,][]{MRW88}:
\begin{equation} 
M_{v}~=~ 189\left(\frac{\Delta v_{1/2}}{km~s^{-1}} \right)^{2} 
\left(\frac{R}{pc} \right), 
\end{equation} 
where R is the radius of the cloud (assumed to be 0.7$\sqrt{ab}$ with
$a$ and $b$ being respectively the FWHM major and minor axes of the
fitted cloud size) and $\Delta v_{1/2}$ is the FWHM of the linewidth.
The constant of proportionality assumes that the GMCs have an
$n\propto r^{-1}$ dependence consistent with what is typically found
for Galactic clouds, but changing the densities anywhere from
$n\propto r^{-(0 - 2)}$ affects the masses by $\lsim$ 50 \%.  While
the spatial resolution is not high enough to precisely determine the
virial mass, it is high enough to provide useful upper limits on the
clouds masses.  Turbulence, stellar winds from the Wolf-Rayet stars
\citep[][]{WR89} and any supernovae associated with the starburst
would all be expected to increase linewidths over their virial value,
so the GMC masses we derive from assuming virial equilbrium should
represent robust upper limits.  Upper limits to the cloud masses based
on virial equilibrium range from $< 8\times 10^{5} M_{\odot}$ (GMC~B)
to $< 1.5\times 10^{7} M_{\odot}$ (GMC~A; Table 2).  Derived values 
for GMC~A should be considered especially uncertain because $\sim$ 1/2 
of the GMC lies outside the half-power points of the primary beam.
 
A third independent constraint on gas masses is from the lack of dust
emission in the millimeter continuum.  For a dust temperature of 46 K
\citep{TT86}, and a 1.3 mm dust absorption coefficient of $3.1 \times
10^{-3}~\rm cm^{2}~g^{-1}$ \citep{PHBSRF94}, $M_{dust}~\simeq ~
220~S^{dust}_{mJy}(1.3mm)~D_{Mpc}^{2} ~M_{\odot}$.  Dust continuum
emission at 1.3 mm is not detected (\S 3.4).  The upper limit to the
1.3 mm dust flux present at any location is 7.5 mJy, which equates to
$M_{dust}~<~2.4\times 10^{4}M_{\odot}$.  If a Galactic gas-to-dust
ratio of 100 is applicable to NGC 5253, the molecular mass implied
from the dust is $< 3\times 10^{6}~M_{\odot}$ for each GMC.  Since NGC
5253 has a low metallicity \citep[log(O/H) = 8.16;][]{KSRWR97}, the
dust-to-gas ratio and hence the molecular cloud gas masses may be
higher than this \citep[][]{SSLH92,LF98}.  However, masses estimated
from 350 $\mu$m continuum observations indicate that the gas-to-dust
ratio is not significantly different from the adopted value (Meier et
al.  2002, in preparation).
 
Surprisingly, for the dust lane GMCs B \& C, the masses estimated from
the virial theorem do not differ significantly from the lower limits
estimated from CO emission and a Galactic conversion factor.
Apparently these two GMCs have near-Galactic conversion factors.  One
would not expect the Galactic conversion to hold in such a low
metallicity galaxy.  To further test whether a Galactic conversion
factor mass is applicable for the dust lane locations, we consider the
relationship between gas mass and $A_{V}$.  Towards the dust lane,
the dust geometry is reasonable well approximated by a foreground
screen, with an $A_{V}~\gsim ~2.2$ \citep[][]{CMBGKLS97}.  
With the standard relationship of N(H)/A$_{V}$ =
$(N_{HI}+2N_{H_{2}})/A_{V}\simeq 1.87 \times 10^{21} ~cm^{-2}
mag^{-1}$ \citep*[][]{BSD78}, this A$_{V}$ implies a column
density of $\gsim 4.1 \times 10^{21}~cm^{-2}$.  The observed HI column
density of $N_{HI}\simeq 2.6 \times 10^{21}~cm^{-2}$ \citep[][]{KS95},
suggests that $N_{H_{2}} \sim 8 \times 10^{20}~cm^{-2}$.  Again, this
value is only slightly less than the $N_{H_{2}} \simeq 1.5\times
10^{21}~cm^{-2}$ obtained from the standard Galactic conversion
factor.  

If an extinction relationship more suitable for NGC~5253's low
metallicity, which is intermediate between those of the LMC and the
SMC, is adopted, N(H) $\simeq ~ 1 \times 10^{22} cm^{-2}~
mag^{-1}$A$_{V}$ \citep*[eg.,][]{WD01}, we obtain column densities of
N(H$_{2}$) $\simeq ~ 10^{22}$ cm$^{-2}$ (at this column density, the
gas would be dominated by $\rm H_2$.)  This value of N(H$_{2}$) is
more than an order of magnitude larger than that found from the a
Galactic conversion factor and more than a factor of two larger than
the upper limit set by the virial analysis.  The assumption of a
Magellenic-like extinction relation significantly overpredicts the
amount of gas present.  Moreover, for $A_{V}\gsim 2$ the
H$\alpha$/H$\beta$ ratio used to derive the visual extinction begins
to saturate.  If the extinction towards the dust lane is larger than
A$_{V}=2.2$, then the above numbers are even more discrepent.
However, since the N(HI) was estimated from observations with a larger
beam ($63^{''}\times40^{''}$) than the CO data, it is possible that
N(HI) could be underestimated if N(HI) is unresolved.  This could
result in a reduction of the derived N(H$_{2}$), but only in the case
of near-Galactic extinction relationships.  Future high resolution HI
observations will be able to address the impact of this effect
(Kobulnicky 2002, in preparation).

Evidently, the narrow linewidths, combined with relatively 
high extinction and bright CO emission imply that CO is not dramatically 
underluminous along the dust lane, and that it does a reasonable
job of tracing the molecular gas mass.  It is possible that CO is not
so underluminous here because the dust lane, which is remote from the
starburst, is not affected by the intense radiation field of the
young stars.
 
The same cannot be said for the other GMCs.  The virial mass upper
limits of the three remaining GMCs A, D \& E are all much larger than
the conversion factor estimates (factors of 10 - 20).  It is
particularly true towards the central obscured star cluster (GMC D).
Dust extinction measurements prove that there are $A_{V}\gsim 15-35$
mag towards the dominant nebula \citep[][]{KNP89,CMBGKLS97}, although
this dust appears to be completely internal to the $\sim$1 pc
supernebula \citep[][, Turner et al. 2002, in
preparation.]{GTB01,TCMB01}.  Moreover, the linewidth of GMC~D could
well be larger than virial, reflecting turbulent motions associated
with the large collection of OB stars.  Such motions would likely not
be present for the other GMCs.  In summary, these results imply that
the dust lane GMCs B \& C have masses of $\sim 10^{6}~M_{\odot}$.  The
masses of the other GMCs are not well determined but probably are
larger than conversion factor values listed in Table 2, $\gsim$few
$\times 10^{6} M_{\odot}$ each.
  
\subsection{Molecular Gas Kinematics: Gas Infall onto the Galaxy?} 
 
Emission from the molecular clouds is detected from $v_{LSR}\sim$355 -
440 km s$^{-1}$, with the bulk of the CO(2-1) emission at $\sim$400 -
420 km s$^{-1}$, consistent with lower resolution CO and HI data
\citep[][]{TBH97,TKS98,MTCB01}. By contrast, the velocity of the
central super-star cluster/supernebula and surrounding $\rm H\alpha$
disk is $\sim$380 km s$^{-1}$ \citep*[][Turner et
al. 2001]{ATPHPH82,MAG01}.  Only GMC~A is significantly blueshifted
relative to central star cluster ($V_{LSR}\simeq 365$ km s$^{-1}$).
In Figure 2, the Position-Velocity (PV) diagram is displayed for gas
integrated within within 5$^{''}$ (which does not include GMC A) of
the minor axis (pa. = 45$^{o}$).  Recessional velocities tend to
increase as one moves farther away from the center of the galaxy along
the dust lane (GMCs D - B), ranging from $\sim$395 km s$^{-1}$ at the
center to $\sim$430 km s$^{-1}$ at or just beyond the field edge.
This is similar to the ionized gas and HI velocity fields at this
location, though the large beam size of the HI observations and the
extinction of the ionized gas makes it difficult to compare in detail
\citep[][]{MK95,KS95}.  Such a trend would be expected if the dust
lane gas is rotating along the {\it minor} axis \citep[][]{KS95}.  If
the molecular gas associated with the dust lane is in rotation around
the minor axis, then GMC E being on the opposite side of the galaxy
should have blueshifted velocities with respect to the galaxy center
(velocities of $\sim$350 km s$^{-1}$ would be expected).  GMC~E has a
recessional velocity of 425 km s$^{-1}$, redshifted from the center of
the galaxy.  A similar upturn in the H$\alpha$ velocity centroid is
seen \citep[crosses in Figure 2 taken from][]{MK95}. If the velocity
centroid of GMC~E is believable and is a continuation of the dust lane
across the galaxy, then the dust lane kinematics are not consistent
with rotation.  Since the dust lane is clearly in front of the
majority \citep[$\sim$90 \%;][]{CMBGKLS97} of the galaxy, the fact
that the dust lane GMCs are all preferentially redshifted relative to
the star cluster implies they are falling into the galaxy.  GMC~A may
be an exception, though the extinction in the galaxy seen in the
direction of this GMC is much lower; perhaps it is falling in as well,
from the back side of the galaxy.  Though difficult to see in the
displayed stretch, inspection of the HST V-band image (Figure 1) shows
increased extinction towards GMC~E as well, suggesting that it too is
foreground and infalling.  If it is assumed that GMC E is not real, it
is possibly that the dust lane GMCs B, C and possibly D could be
rotating, but then GMC A would still be infalling.  As a result, it is
concluded that over the central $\sim 40^{''}$ radius of the galaxy,
the kinematics of the molecular gas favor infall (accreting gas) over
rotation.  Which gas is infalling, GMCs A \& E or the dust lane GMCs B
\& C, depends on the believability of GMC E, but in any case it
appears from the kinematics that some gas must currently be infalling.
CO traces only part of the extensive cloud cloud of HI gas, so higher
resolution HI observations (Kobulnicky 2002, in preparation) will
give a much more complete picture of the complex gas kinematics in
NGC~5253.
 
\subsection{Millimeter Continuum: Further Evidence for Optically  
Thick Cm-wave Emission from the Supernebula} 
 
The radio continuum spectrum of NGC 5253 is flat from $\sim$20 cm to
2.6 mm \citep*[][]{BTHLK96, TBH97}.  At short wavelengths, the central
``supernebula'' associated with a very young SSC
contributes a significant fraction of the radio flux, but the emission 
is likely
optically thick longward of $\sim$2 cm \citep*[][]{THB98,TBH00}.  Even
the highest frequency cm-wave observations may still not detect the
full strength of the central starburst in NGC~5253.  
Emission from dust can also
potentially contribute to the observed millimeter
continuum fluxes.  Therefore, we have imaged the millimeter continuum
at both 3.1 mm and 1.3 mm to better constrain the spectral energy
distribution, the total star formation strength and the dust content.
 
Figure 3 displays the 97 GHz (3.1 mm) and 233 GHz (1.3) mm continuum
maps of NGC 5253.  The 3.1 mm continuum emission is essentially
unresolved in the $9.^{''}0 \times 6.^{''}8$ beam, with a marginally
believable extension to the west of the peak.  The peak 3.1 mm flux
density of the central source is $34\pm 4$ mJy beam$^{-1}$.  Summed
over the inner $20^{''}$, the total 3.1 mm flux density is $54 \pm 5$
mJy.  This is in good agreement with the values $38 \pm 6$ and $52 \pm
8$ at 2.6 mm \citep*[][]{TBH97}.  As in the case of the 3.1 mm map,
the emission at 1.3 mm is basically unresolved, though it is possible
that there is a western extension as well.  The 1.3 mm peak flux
density is $38 \pm 8$ mJy and $46 \pm 10$ over the central 20$^{''}$.
The 1.3 mm continuum peak is centered at $\alpha = 13^{h} 39^{m}
56.15^{s}$, $\delta = -31^{o} 38^{'} 24.^{''}6$.  The 3.1 mm continuum
peak is $\sim2^{''}$ west of the 1.3 mm position, but considering the
large beams and moderate SNR, this difference is not significant.  We
are unable to confirm the existence of the suggested weak secondary
northern component \citep[][]{TBH97}.  The spectral index, $\alpha$,
($S_{\nu}\propto \nu^{ \alpha}$) found for the central cluster between
3.1 mm and 1.3 mm is $\alpha^{3.1}_{1.3}\simeq -0.18\pm0.30$.  The
flat spectral index demonstrates that the radio emission from the
central starburst is dominated by thermal bremsstrahlung emission out
to 230 GHz (see below).
 
In Figure 4, the radio-mm spectral energy distribution (SED) is
displayed for the central 20$^{''}$ of NGC 5253.  The additional flux
measurements are taken from \citet[][]{TBH97} and \citet[][]{THB98}.
The radio spectrum is indeed basically flat from 20 cm to 1.3 mm.
However, interesting deviations from a flat ($\alpha = \nu^{-0.1}$)
spectrum are seen.  The total continuum fluxes observed at the
shortest wavelengths ($<$1 cm) are brighter than extrapolated from the
centimeter data assuming the standard optically thin bremsstrahlung
spectrum.  The probable explanation for this, as suggested previously,
is optically thick radio continuum from the supernebula associated
with the youngest optically obscured cluster \citep[][]{THB98,TBH00}.
 
We model the radio-mm SED based on the thermal/non-thermal separation
derived from the cm-wave data \citep[][]{THB98}.  The SED is fit with
three components.  One component is optically thin, thermal free-free
emission with a flux density of 43 mJy at 6 cm, based on summing up
the 6 cm flux of the sources with thermal spectral indices between 2
cm and 6 cm.  The second component is synchrotron emission with an
adopted spectral index of $\alpha = -0.75$, which accounts for the
remainder of the flux at 20 cm.  This component is very small ($\sim$8
mJy) and negligible shortward of 10 cm.  The third component is the
supernebula.  It is assumed to have an optically thick free-free
spectrum with a flux density of 11 mJy at 2 cm and a turnover
wavelength ($\lambda_{o}(\tau=1)$) below 2 cm as seen in the VLA-A
array map \citep[][]{TBH00}.  That the millimeter fluxes are only
slightly brighter than the 2 cm point demonstrates that the turnover
wavelength cannot be much shorter than 1.3 cm.  The best fit to the
turnover wavelength is $\sim 1.5\pm0.2$ cm (20 GHz).  Therefore, the
higher flux densities seen at the millimeter wavelengths beautifully
confirm the presence of an HII region that is optically thick below 2
cm.  While it is possible that NGC 5253 contains several optically
thick HII regions with different turnover wavelengths, the quality of
the fit without them indicates that any additional sources must not
provide a substantial contribution to the overall flux.  Again, NGC
5253 is unique, having a single, optically thick supernebula that is
dominant enough to be distinguishable in the global SED of the
nucleus.
 
From the spectrum and $\alpha^{3.1}_{1.3}$ it is clear that the
continuum detected at 1.3 mm is still consistent with being entirely
thermal bremsstrahlung.  No dust emission is detected at 1.3 mm.  From
the 1.3 mm continuum flux, we estimate \citep[eg.,][]{MH67} a
Lyman continuum rate of $6 \times 10^{52}~s^{-1}$ for the central
20$^{''}$, \citep[for $1.2\times10^{4}$ K;][]{WR89}.  The rate is
somewhat higher than estimated at 2 cm, as expected if some of the
free-free emission is still optically thick at that wavelength.  From
the SED (Figure 4) at least N$_{Lyc}~ \gsim 3 \times 10^{52}~s^{-1}$
comes from the optically thick supernebula alone.  The $N_{Lyc}$ rate
derived from the 1.3 mm continuum data should be a robust estimate of
all possible ionizing photons in the central region, since an emission
measure, $EM~>~10^{11}~cm^{-6}pc$, would be required for optically
thick continuum beyond 200 GHz.  Even in an exotic region like NGC
5253's supernebula, this is unlikely for this relatively large
region.  Given that the best fit
turnover wavelength is $\sim 1.5$ cm, the central supernebula has an
$EM \simeq 2.1 \times 10^{9} cm^{-6} pc$, or $<n_{e}>^{1/2}\simeq 4.6
\times 10^{4} cm^{-3}$ for a line-of-sight distance of 1 pc.  Towards
the dust lane, an upper limit of $N_{Lyc}~<~3 \times 10^{51}~s^{-1}$
is derived from the more constraining 3.1 mm continuum upper limit.
Apparently, the dust lane is not a site of strong current star
formation despite being the brightest region in CO.

\section{Molecular Gas and the Star Formation Efficiency in NGC 5253} 
 
The confinement of the molecular gas to a lane crossing the minor
axis, together with the fact that the foreground molecular gas is
redshifted with respect to the starburst strongly suggests that gas is
currently falling into the nucleus.  The idea that an interaction with
the ``companion'', M~83, (at a current projected distance of 130 kpc)
occurred $\sim$1 Gyr ago, triggering the burst of star formation, has
been suggested by several authors
\citep[][]{vB80,G81,KS95,TBH97,CCGK99}.  In a global sense, this is a
reasonable trigger for the starburst in NGC 5253.  However, due to the
large differences in timescales, $\sim$1 Gyr for the interaction age
and $<10$ Myr for the current starburst episode, the current burst of
star formation can only be indirectly triggered by an interaction with
M~83.  No other suitable galactic merger candidates appear near enough
to NGC 5253 to directly trigger the burst.  We find a similar time
delay between the interaction and the current starburst in another
nearby dwarf interaction-induced starburst, NGC 3077
\citep[][]{MTB01}.  In that starburst, we demonstrated that the delay
can successfully be explained by the gas clouds that were either
pulled out of the dwarf, NGC 3077, or captured from M 81 raining back
down onto the galaxy due to effects such as dynamical friction.
Evidence is strong for the triggering in NGC 3077; in this famous
system (the M81-M82-NGC 3077 triplet), large streamers of gas connect
the three main galaxies.  In addition, large gas clouds are seen just
outside the optical disk of NGC 3077, and the starburst in the galaxy
is seen at the interface of two ``colliding'' GMCs.
 
It is tempting to apply the same mechanism to NGC 5253, although
direct evidence for an interaction with M 83 is not available.
Indirect support of such an interaction is provided by the strongly
warped disk of M83, a finger of HI emission extending towards NGC 5253
\citep[][]{RLW74,TA93,CTBHM02}, the fact that the dust lane points
back towards M83 \citep[eg.,][]{CCGK99} and the existence of a
population of intermediate age clusters in the outer parts of NGC 5253
\citep[between 0.1 - 1 Gyr;][]{CP89}.  On the other hand, low
resolution HI observations of NGC 5253 do not show evidence for any HI
clouds greater than $10^{6}$ M$_{\odot}$ outside its optical disk
\citep[][]{KS95}, unlike what has been seen in the M81 group
\cite[][]{YHL94}.  

Clearly though, the presence of infalling gas along the minor axis (\S
3.3), implies that some gas must have been in the halo sometime in the
recent past.  In fact, from a conceptual perspective, infall of a gas
cloud into a dwarf galaxy model may represent a qualitatively similar
event to the large spiral-satellite mergers modeled, only on a much
smaller sizescale, with the delayed trigger of the starburst occurring
upon ``the final stages'' of merger of the small gas cloud
\citep[eg.,][]{H89,MH94}.  Assuming the conversion factor over much of
the galaxy is not significantly larger than we have estimated (\S
3.2), a $10^{6}$ M$_{\odot}$ cloud is all that is required to explain
the dust lane molecular gas.

The origin of the gas in the dust lane is unknown.  NGC~5253 appears
to have been a metal-poor dwarf elliptical before the current burst
with very little galactic rotation \citep[][]{CP89}. This suggests
that the bulk of the infalling gas did not originally belong to
NGC~5253.  In fact, if the conversion factors we derived in \S 3.2 are
correct, the dust lane gas may have higher metallicities than the rest
of the ``disk'' of NGC~5253.  This situation could arise if the dust
lane gas came from a higher metallicity object like M~83.  It might be
expected that the gas morphology seen in NGC~5253 is similar to what
NGC~3077 will look like when most of its remaining halo gas clouds
settle into the galaxy, a few hundred Myrs from now.  In order to
conclude that the gas originated from NGC 5253, it appears that one
would require the gas to experience enrichment in metals from a
previous generation star formation event as it falls back in from the
halo \citep[eg.,][]{MF99, DB99, RMD01}.
 
The molecular gas morphology and the extreme youth of the starburst in
the NGC 5253 put strong constraints on the gas evolution in the
nucleus.  The youngest optically visible cluster is 2.5 Myrs old
\citep[][]{CMBGKLS97}, and the optically obscured super-star cluster
and its associated supernebula are almost certainly even younger,
$\sim 1$ Myr or less \citep[eg.][]{BTHLK96,CBWCMS99,TBH00}.  Assuming
that the GMCs are moving $\lsim$ 40 km s$^{-1}$ with respect to the
super star cluster/supernebula (\S 3.3), only GMCs within a distances
of $\delta R \sim ~\delta v \times \tau_{burst}~<~ 120$ pc
(6.$^{''}$5), can be casually connected with it.  Of the molecular gas
detected only GMC~D can satisfy such a constraint.  The molecular gas
we detect along the dust lane is too far away and moving in the wrong
direction to be responsible for the current burst of star formation.
Presumably it is the fuel of a future burst.

It is well known that the IR luminosity of NGC 5253 is high relative
to the amount of molecular gas present, $L_{IR}/M_{H2} \sim 800~
L_{\odot}/M_{\odot}$ \citep[][]{TBH97}, suggesting a high global
efficiency of star formation.  We consider here the SFE local to
within $\sim 100$ pc of the SSC.  Based on a Salpeter IMF,
\citet[][]{TBH00} calculate that the central cluster contains $\sim
10^{6} ~M_{\odot}$ of stars.  Assuming that only the molecular gas in
GMC D can be associated with the current SSC formation, this implies
$M_{*}/M_{H_{2}} \sim 3$, for a Galactic conversion factor, or a SFE =
$M_{*}/(M_{*}+M_{H_{2}})$ = 75 \%.  Even if the applicable conversion
factor for NGC 5253 is a factor of 10 larger than the Galactic value,
the upper limit allowable from the virial linewidth (which probably
overestimates the molecular mass somewhat; \S 3.2), the SFE on this
localized scale is quite high, at least 25 \%.  The SFE will be even
higher if GMC D is not directly related to the burst, being just lined
up in projection.  Apparently, this super-star cluster has converted a
major fraction of its gas mass to stars, as expected in order for the
cluster to remain bound \cite[eg.][]{LMD84}.

One of the goals of this project is to clarify what triggered the
formation of the young super star clusters in NGC~5253.  Because the
current starburst is so young, a few Myrs, the molecular gas
distribution should not have had time to change significantly since
its ``initial configuration''.  The current starburst appears to occur
at the ``intersection'' of the dust lane and the center of the galaxy.
This morphology supports the hypothesis that the burst of star
formation in NGC~5253 is being triggered by inflow of gas. The
comparative lack of star formation associated with dust lane itself
(\S3.4), implies that strong star formation does not begin until the
infalling gas reaches the center of the galaxy.  A possible
explanation for this could be that there existed a collection of gas
(possibly HI) at the center of NGC 5253's potential well that the
infalling gas collided with.  If so, the burst would have had to have
consumed its gas extremely efficiently in order to create several
$\sim10^{5}-10^{6}~M_{\odot}$ clusters \citep []{CMBGKLS97,THB98} with
only the extremely faint GMC~D remaining of the gas.  One can relax
the efficiency somewhat by assuming that the CO conversion factor at
GMC~D is off by orders of magnitude \citep[][]{TBH97}; however our
analysis indicates that the conversion factor cannot be large enough
to eliminate the need for a high star formation efficiency.  Another
possible trigger is that the GMCs shock and collapse due to the
steepening of the galaxy potential at the center of the nucleus
\citep[eg.,][]{KL92,W93}.  This is unlikely for a small galaxy like
NGC 5253.  A third possibility is that an overpressure on the GMC
surfaces due to a superwind driven from the older clusters in the
nucleus of NGC 5253 \citep[eg.][]{G81,MHWS95,MK95,SS99} compressed the
infalling clouds as they approach the center of the galaxy, triggering
the current burst, a form of sequential star formation \citep{EL77}.
This is an enticing hypothesis since the ages of the brightest star
cluster do appear to decrease consistently from the southwest towards
the northeast hinting at the possiblity of propagating star formation
\citep[][]{G96,CMBGKLS97,TCLH01}. However, evidence for a superwind is
weak. All observed linewidths to date have been relatively narrow
\citep{MHWS95,MAG01,T02}. ROSAT HRI observations of the X-ray emitting
gas imply that the hot gas is still localized to the star clusters and
hasn't yet had time to drive a large-scale galactic wind
\citep[][]{SS99}.  And this theory is not obviously consistent with
the current distribution of molecular gas: why it there an age trend
southwest-to-northwest, when the gas is distributed
southeast-to-northwest? Super star clusters, which require a high
efficiency of conversion of gas into stars, may leave little evidence
of their formation in the form of molecular gas.

\section{Summary}

We have confirmed the detection of CO in the dust lane of NGC 5253.
Our analysis of the dust and virial masses indicate that the Galactic
conversion factor may hold locally in the dust lane, implying it
probably contains less than a few $\times 10^{6}~M_{\odot}$ of
molecular gas.  For the first time we detect molecular gas in the
direction of the central SSCs and radio/IR nebula; the total amount of
molecular gas at this location is small, $\sim
10^{5}$--$10^{6}M_{\odot}$. We also confirm the presence of optically
thick free-free emission at 2~cm towards the supernebula, and obtain
N$_{Lyc} \simeq 6 \times 10^{52} ~s^{-1}$ over the central 20$^{''}$.
The kinematics of the molecular clouds in the dust lane indicate that
these clouds are falling into NGC~5253, consistent with the hypothesis
that the starburst is caused by accretion of gas from outside the
galaxy.  However beyond this general inference, little can be said
about the detailed trigger for the star formation. The small amount of
molecular gas close enough to be causally associated with the current
star formation demonstrates that regardless of what the specific
trigger is it must have been extremely efficient.  We estimate a star
formation efficiency of $\sim$75 \% for the current starburst: at
worst it can be $\sim$ 25 \%.  This high efficiency is consistent with
the formation of a large, bound cluster as indicated by the radio/IR
supernebula. The gun that triggers super star cluster formation may be
relatively smokeless.

\acknowledgements 
 
We are grateful to the faculty, staff and postdocs at OVRO for their
support and assistance during the observations.  We thank Varojan
Gorjian and Chip Kobulnicky for helpful discussions and Daniella
Calzetti for the use of the HST images of NGC 5253.  We also thank the
referee for comments that help in clarifying and focusing the paper.
This work is supported in part by NSF grant AST-0071276 to J.L.T. and
by a grant from the Israel Academy Center for Multi-Wavelength
Astronomy to S.C.B.  The Owens Valley Millimeter Interferometer is
operated by Caltech with support from the NSF under Grant 9981546.
This research has made use of the NASA/IPAC Extragalactic Database
(NED) which is operated by the JPL, Caltech, under contract with the
National Aeronautic and Space Administration.

\clearpage 

\begin{deluxetable}{lccccc} 
\tablenum{1} 
\tablewidth{0pt} 
\tablecaption{Observational Data\tablenotemark{a}\label{tab1}} 
\tablehead{ 
\colhead{Transition\tablenotemark{b}}  
&\colhead{Frequency}  
&\colhead{$\Delta V_{chan}$} 
&\colhead{$\Delta \nu_{band}$}  
& \colhead{Beamsize}  
& \colhead{Noise level} \\ 
\colhead{}  
&\colhead{\it (GHz)}  
&\colhead{($km~s^{-1}$)} 
&\colhead{\it (MHz)} 
& \colhead{\it (arcsec; deg)}  
& \colhead{\it (mK / mJy Bm$^{-1}$)}}  
\startdata 
CO(2-1)& 230.538 & 5.20\tablenotemark{c}& 32 &$9.6\times 4.8;-84^{o}$ 
\tablenotemark{d} & 50/0.10\\ 
3.1 mm & 96.5  &\nodata  &1000 &$9.0\times 6.8;19^{o}$ & 5.4/2.5\\ 
1.3 mm  & 233.3  &\nodata  &1000 &$6.5\times 4.5;-86^{o}$ & 5.9/7.5\\ 
\enddata 
\tablenotetext{a}{For observations made from 1996 November 16 - 1997  
December 21.} 
\tablenotetext{b}{Phase Center \#1: $\alpha = 13^{h} 37^{m} 04^{s}.8~~ 
\delta = -31^{o} 22' 58.^{''}0$ (B1950)\\ 
$~~~~~$Phase Center \#2: $\alpha = 13^{h} 37^{m} 05^{s}.6~~ 
\delta = -31^{o} 23' 22.^{''}0$ (B1950)} 
\tablenotetext{c}{The CO(2-1) maps were smoothed to 5.2 km s$^{-1}$  
resolution from 2.6 km s$^{-1}$.} 
\tablenotetext{d}{The CO(2-1) data was tapered with a 40k$\lambda$ taper.} 
\end{deluxetable} 
 
\clearpage 
 
\begin{deluxetable}{lcccccc} 
\tablenum{2} 
\tablewidth{0pt} 
\tablecaption{Giant Molecular Clouds in NGC 5253\label{tab2}} 
\tablehead{ 
\colhead{Parameter} 
&\colhead{Units} 
&\colhead{}  
&\colhead{} 
&\colhead{GMCs} 
&\colhead{} 
&\colhead{} 
\\ 
\colhead{}   
&\colhead{} 
&\colhead{A\tablenotemark{a}} 
&\colhead{B} 
&\colhead{C} 
&\colhead{D} 
&\colhead{E} 
} 
\startdata 
Right Ascension\tablenotemark{b}  & $13^{h}39^{m}$  
& $\gsim 57.^{s}3$ &$57.^{s}2$  &$56.^{s}6$  & $56.^{s}0$ &$55.^{s}5$  \\ 
Declination\tablenotemark{b}  &$-31^{o}38^{'}$   
&30$^{''}$  &36$^{''}$  &32$^{''}$  &26$^{''}$  &18$^{''}$  \\ 
$a \times b$\tablenotemark{b,c} & $pc \times pc$ 
&$<89$  &$<89$  &$\lsim 89$  &$\sim 220 \times 110$  &$<89$   \\ 
$v_{o}$\tablenotemark{b}  & $km ~s^{-1}$  
&363  &421  & 419  &395  & 424  \\ 
$\Delta v_{1/2}$\tablenotemark{b}  & $km ~s^{-1}$  
&36  & 8.4 & 12 & 15 &27  \\ 
$I_{pk}$\tablenotemark{b}  &Jy   
&0.36  & 1.0 &0.90  &0.51  &0.30  \\ 
$S_{CO}$\tablenotemark{d}  & $Jy~km ~s^{-1}$  
&16.9  &12.7  &14.3  & 6.9 & 7.6 \\ 
$M_{CO}\left(\frac{X_{CO}}{2.0 \times 10^{20}} \right)$  & ($\times 
10^{6}$) $M_{\odot}$  
&0.86  &0.65  & 0.73 &0.36  &0.38  \\ 
$M_{vir}\tablenotemark{e}$  &($\times 10^{6}$) $M_{\odot}$   
&$<15$  &$<0.83$  &$\lsim 1.7$  &$\sim$4.6  & $<8.5$ \\ 
\enddata 
\tablecomments{Uncertainties are conservately estimated as $\pm
2^{''}$ for the position centroids, $\pm 4.^{''}$8 for cloud sizes, 3
km s$^{-1}$ for the linewidth and line centroid, 0.1 Jy for the peak
intensity, 3 Jy km s$^{-1}$ for the flux density and approximately a
factor of two for the masses.}
\tablenotetext{a}{Fitted values for GMC A are more uncertain given
that it is at the edge of the primary beam FWHM.  Integrated intensities 
of this cloud make no attempt to correct for emission outside the 
field of view.}
\tablenotetext{b}{Values based on gaussian fits to the spectra.} 
\tablenotetext{c}{Deconvolved sizes smaller than the beam minor axis  
($4.^{''}8$ = 89 pc) are considered unresolved.} 
\tablenotetext{d}{Flux estimates are based on the assumption that the  
clouds are unresolved.  The applied 1$\sigma$ clipping (\S 2) may cause 
a slight underestimate of cloud sizes, but fitting the GMCs with 
no clipping demonstrates that this does not increase any derived 
sizes greater than the quoted upper limits.} 
\tablenotetext{e}{Virial masses are estimated assuming a cloud radius of  
0.7$\sqrt{ab}$ \citep[eg.][]{MT01}} 
\end{deluxetable} 

\clearpage 

\begin{figure}
\epsscale{1.0}
\plotone{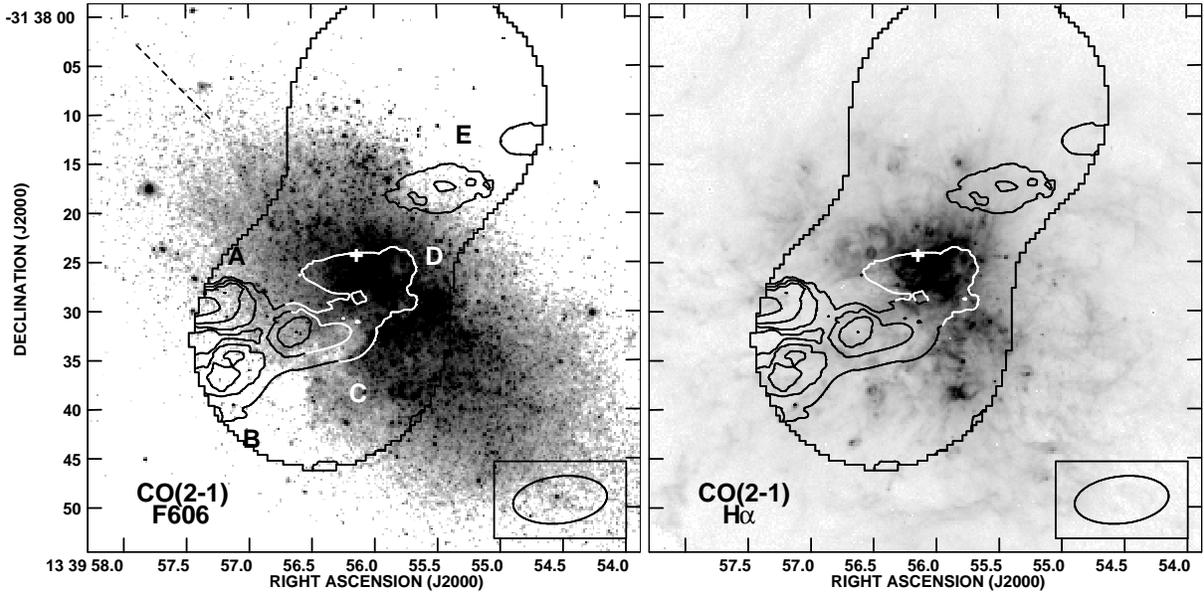}
\caption{The CO(2-1) integrated intensity overlaid
on the a) F606W HST image and b) the HST H$\alpha$
\citep[]{CMBGKLS97}.  Contours are in steps of 1.8 K km s$^{-1}$
($\sim 3 \sigma$).  Letters label the five GMCs.  The cross marks the
location of the centroid of the 1.3 mm continuum.  The figure-eight
shape curve marks the FWHM power points of the field-of-view.  The
beam is plotted in the lower right.  the dashed line represents the
adopted P.A. = 45$^{o}$ (or 225$^{o}$).  Given the small
field-of-view, uncertianties in astrometry (particularly field
rotation) of the HST image relative to the millimeter are possible at
the $\sim 2^{''}$ level.}
\end{figure}

\clearpage 

\begin{figure}
\epsscale{0.6}
\plotone{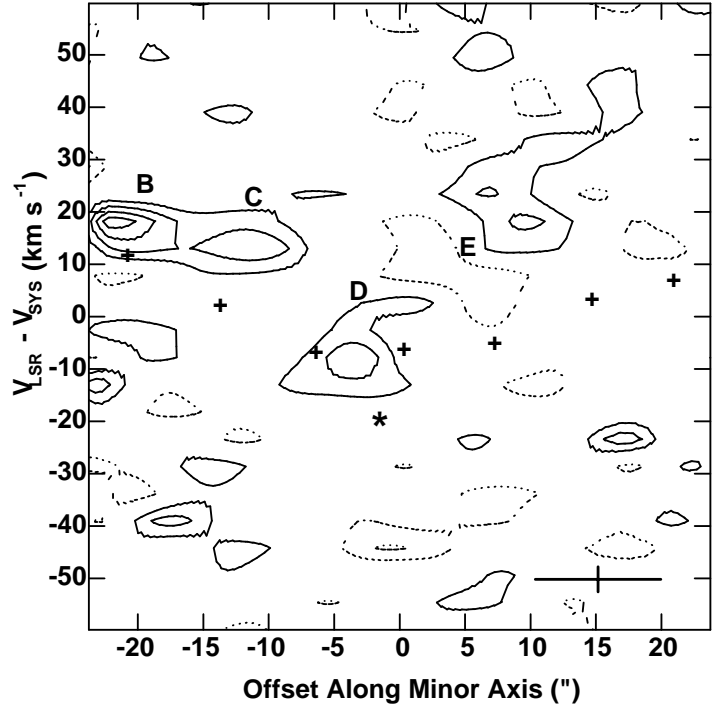}
\caption{Position-Velocity diagram for NGC 5253
integrated over gas within 5$^{''}$ of the {\it minor} axis (P.A. =
45$^{o}$).  Zero velocity corresponds to a $v_{LSR}= 405$ km s$^{-1}$,
and the zero position corresponds to $\alpha = 13^{h}39^{m}56.1^{s};
~\delta=-31^{o}38^{'}22.^{''}5$.  The GMCs are labeled.  (Much of GMC
A is not within 5$^{''}$ of the minor axis and hence does not feature
prominantly in the figure.  A small fraction of emission from GMC A
can be seen contributing at the feature seen at $v\simeq -40$ km
s$^{-1}$ and -18$^{''}$.)  The crosses mark the velocity of the
h$\alpha$ gas measured along the same position angle \citep[]{MK95}.
The position and velocity of the super-star cluster is marked with an
asterisk \citep[][, Turner et al. 2002, in preparation]{TBH00,MAG01}.
The characteristic resolutions for each axis are marked in the lower
right.}
\end{figure}
 
\clearpage 

\begin{figure}
\epsscale{1.0}
\plotone{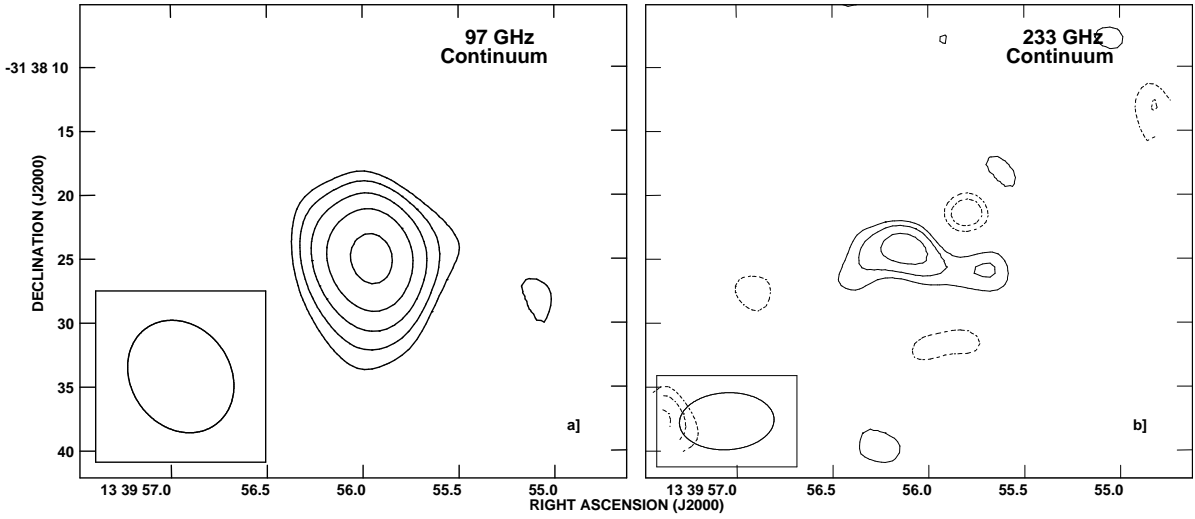}
\caption{Millimeter continuum in NGC 5253. a) The 97 GHz  
(3.1 mm) continuum map.  Contours are in steps of $\pm 2^{n/2}$, n=0,1, 
2... times 7.5 mJy beam$^{-1}$ (3$\sigma$), with dashed contours negative.  
b) The 233 GHz continuum map.  Contours are in steps of $\pm 2^{n/2}$, n=0,1, 
2... times 15 mJy beam$^{-1}$ (2$\sigma$), with dashed contours negative.   
Beamsizes are indicated at the bottom left of each map.} 
\end{figure} 

\clearpage 

\begin{figure}
\epsscale{0.8}
\plotone{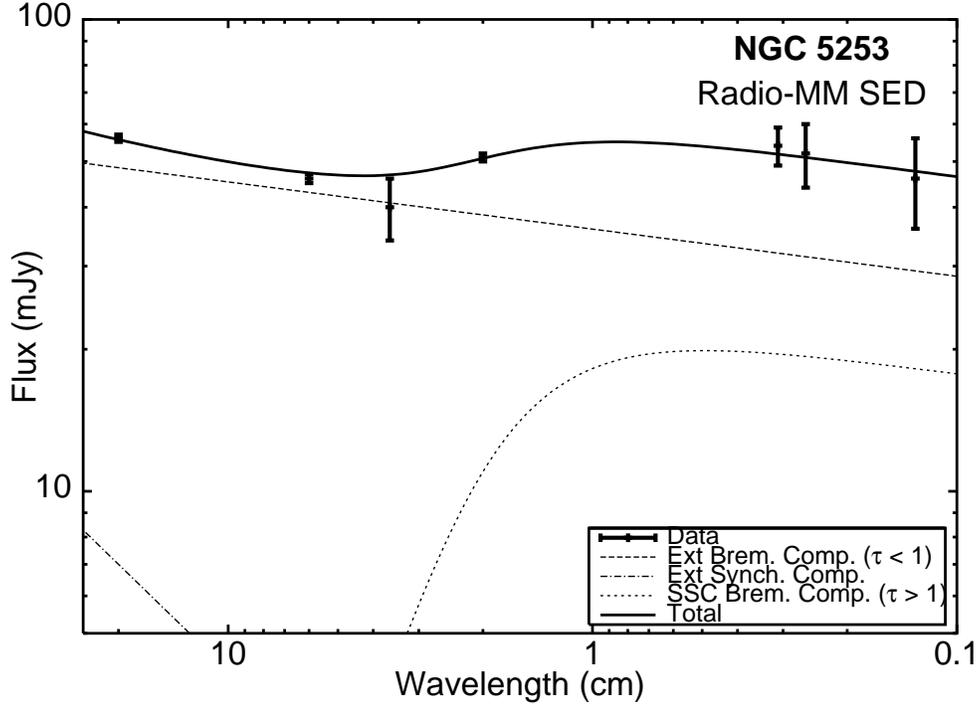}
\caption{The radio / millimeter spectral energy
distribution of NGC 5253.  The 20 cm - 1.3 cm data are taken from
\citep[][]{THB98}, and the 2.6 mm datapoint is from \citep[][]{TBH97}.
Fits are described in the text.  The dotted line represents the
optically thick young super-star cluster component that contributes 11
mJy flux at 2 cm and has a turnover wavelength of 1.5 cm.  The dashed
line represents the extended thermal free-free emission with $\alpha =
-0.1$ and a flux that matches the thermal 6 cm flux (43 mJy).  The
dot-dashed line is the synchrotron component ($\alpha = -0.75$)
predicted based on matching to the observed 20 cm flux
\citep[][]{THB98}.  The solid line is the combined fit to the spectral
energy distribution.  Note that the 3.6 cm datapoint should be treated
as an lower limit since it is not from matched VLA array
observations.}
\end{figure} 
 
\end{document}